\documentstyle[preprint,aps]{revtex}

\def\slashchar#1{\setbox0=\hbox{$#1$}  
   \dimen0=\wd0     
   \setbox1=\hbox{/} \dimen1=\wd1  
   \ifdim\dimen0>\dimen1   
      \rlap{\hbox to \dimen0{\hfil/\hfil}} 
      #1     
   \else     
      \rlap{\hbox to \dimen1{\hfil$#1$\hfil}} 
      /      
   \fi}      %
\def\mbf#1{\mbox{\boldmath${#1}$}}

\begin{document}

\draft
\preprint{\parbox[t]{3.5in}{\begin{flushright}
                            RUB-TPII-55/93, TPR-93-38, November 1993\\
                                                      hep-ph/9312219\\
                             \end{flushright} }}

\title{\LARGE \bf Electric Polarizability of the Nucleon\\
in the Nambu--Jona-Lasinio  Model}

\author{\large Emil N. Nikolov$^{1}$\footnote
{\parbox[t]{6.34in}{\baselineskip6mm On leave of absence from
Institute for Nuclear Research and Nuclear Energy, 1784 Sofia, Bulgaria;\\
DAAD fellow}}
 ,
 Wojciech Broniowski$^{2}$\footnote
{\parbox[t]{6in}{\baselineskip6mm On leave of absence from
   Institute of Nuclear Physics, PL-31342 Cracow, Poland;\\
Alexander von Humboldt fellow}}
\setcounter{footnote}{0}\footnotetext{E-mail addresses: \parbox[t]{3.5in}
{\baselineskip6mm emiln {\em or} goeke@hadron.tp2.ruhr-uni-bochum.de\\
broniowski@rphs1.physik.uni-regensburg.de}}
,
 and Klaus Goeke$^1$}

\address{\sl $^1$ Institut f\"ur Theoretische Physik II,
Ruhr-Universit\"at Bochum\\ 
D-44780 Bochum, Germany}

\address{\sl $^2$ Institut f\"ur Theoretische Physik II,
Universit\"at Regensburg\\
D-93040 Regensburg, Germany}

\maketitle

\begin{abstract}
The electric polarizability of the nucleon is calculated in the soliton
approach to the Nambu--Jona-Lasinio model. We analyze the leading-$N_c$ 
contributions, as well as the effects of rotational $1/N_c$ corrections 
and $\Delta$-$N$ mass splitting. Our model prediction is substantially 
reduced compared to other soliton calculations, and is closer to the 
experimental value.
\end{abstract}


\section{Introduction} \label{se:intro}
Electromagnetic polarizabilities are fundamental properties
of hadrons which are manifest in 
various processes involving two photons
\cite{Petrunkin81,Friar:rev}.
Recent measurements 
\cite{Federspiel,Zieger,Schmiedmayer} of 
the electric, $\alpha$, and magnetic, $\beta$, polarizabilities
of the nucleon narrowed considerably experimental uncertainties in
these observables, 
and were accompanied by a number of theoretical studies. 
Attempts to
describe $\alpha$ and $\beta$ were made in various approaches, ranging from 
chiral perturbation theory
\cite{BKM91,Butler} and dispersion relations \cite{Lvov:disp}
to chiral soliton models 
\cite{Chemtob87,Scoccola90,BBC91a,polar,Schwesinger:pol,Golli:pol}.
Earlier calculations of $\alpha$ and $\beta$ are reviewed in 
Ref.~\cite{DMP1}.

In this paper we calculate the average electric polarizability 
of the proton and neutron, 
\mbox{$\alpha \equiv \frac{1}{2} (\alpha_p + \alpha_n)$},
using the solitonic approach \cite{Meissner89,Reinhardt88} to the 
Nambu--Jona-Lasinio model \cite{NJL}. 
This  model has led to 
quite successful phenomenology \cite{Meissner91,Christo:ff}.
In particular, baryonic mass relations, magnetic moments, 
and various form factors are reproduced well within
expectations. It is therefore challenging to see if we can
also describe {\em two-current observables}, such as the polarizability, 
in the NJL model. 

There are two important features in our study which 
have not been considered in earlier works in soliton models: 
1)~the role of the Dirac sea effects (this can only be done in a model
which has the Dirac sea, such as the NJL model), and
2)~inclusion of rotational
$1/N_c$-effects.
Such effects were recently analyzed for the case of the axial coupling 
constant $g_A$ in Ref.~\cite{Wakamatsu93}
and were more closely examined in Refs.~\cite{Christo:ga,Michal:ga}.

In the NJL model the only dynamical 
degrees of freedom are quarks, which
occupy valence orbitals, as well as the Dirac sea. 
We find that the effects of the Dirac sea in our calculation
of $\alpha$ are very well reproduced
by the first two terms in the gradient expansion, where the first
term is
the so called ``seagull'' contribution to $\alpha$,
discussed in many previous papers \cite{Chemtob87,Scoccola90,polar}.
Our analysis 
demonstrates explicitly that the inclusion of the sea-gull in the
$\sigma$-model or in the Skyrme model does not
violate gauge invariance, as has recently been claimed in 
Ref.~\cite{Lvov:gauge}.
This is readily seen from the full NJL expression, which is purely
{\em dispersive} in character 
({\em i.e.} involves no seagull terms) 
and manifestly gauge invariant. The seagull
term emerges from the full expression when the gradient expansion is
performed 
(see Sec.~\ref{se:leading}). 

The inclusion of rotational $1/N_c$ effects has very sound phenomenological 
consequences. As discussed extensively in Ref.~\cite{polar}, 
chiral soliton models, when treated at the leading-$N_c$ level, 
have problems in predicting electromagnetic polarizabilities
correctly. The dominant 
contribution to the electric polarizability is
obtained from pion tail effects, and is proportional
to $g_A^2$. If one insists that the model fits the value
of $g_A$ (which one should!), then the value of $\alpha$ obtained
at the leading $N_c$ level is roughly a factor of $2 - 3$ too large,
as demonstrated in the $\sigma$-model calculation of Ref.~\cite{polar}.
In the NJL model (as well as in the Skyrmion) the value of 
$g_A$ obtained with various fits has always been notoriously
too small at the leading-$N_c$ level \cite{Meissner91:ga}, {\em i.e.} 
of the order $0.7 - 0.8$. 
As first noticed in Ref.~\cite{Wakamatsu93}, the inclusion  
of rotational $1/N_c$ corrections in the NJL model is capable of raising  
$g_A$ to a comfortable value of $\sim 1.2$ \cite{Christo:ga,Michal:ga}. 
This is a big correction: 
it raises $g_A^2$ by a factor of $2-3$. 
We calculate analogous $1/N_c$ corrections for the electric polarizability,
and find a sizeable contribution. 
As a result, and
after including approximately
additional corrections due to the $N$-$\Delta$ mass splitting
\cite{chihog}, we are able to obtain
a number which is closer to experiment than in other studies
in soliton models
\cite{polar,Golli:pol}, but still  too large.
For the typical choice \cite{Meissner91,Christo:ff}
of the constituent quark mass \mbox{$M = 420\
\rm{MeV} $} 
 we obtain
\mbox{$\alpha \simeq 19 \times 10^{-4}\ {\rm fm}^3$}
compared to the experimental value
$\alpha_{\rm exp}=9.6 \pm 1.8 \pm 2.2
\times 10^{-4}\ \rm fm^3$ \cite{Federspiel,Zieger,Schmiedmayer}.

In this paper we do not analyze the splitting of the proton and neutron
polarizabilities, since it involves a complicated problem of 
treating the translational zero mode \cite{BBC91a}. Also, the magnetic
polarizability is not analyzed. It was shown in Ref.~\cite{polar} that
the large-$N_c$ approach is not a good starting point to describe this
observable. Hence, we concentrate solely on the average proton and
neutron electric polarizability, 
$\alpha$.

The outline of the paper is as follows: In Sec.~\ref{se:form} we
develop the necessary formalism of linear response theory for the 
NJL model, and derive a basic expression for the electric
polarizability. In Sec.~\ref{se:leading} we calculate the 
leading-$N_c$ contribution to $\alpha$, which comes from both
valence and sea quark effects. 
In Sec.~\ref{se:subleading} we calculate the rotational $1/N_c$
corrections to $\alpha$. The valence contribution is calculated exactly, 
while the sea part is estimated using gradient expansion. 
Additional corrections, due to $\Delta$-$N$ mass splitting, are
discussed in Sec.~\ref{se:splitting}. Sec.~\ref{se:conclusion}
contains our results and conclusions.

\section{Formalism} \label{se:form}
Polarizabilities are defined as coefficients in the low-energy
expansion of the Compton amplitude \cite{Petrunkin81,Friar:rev,Lvov:gauge}.
It has been shown 
\cite{Petrunkin81,Friar:rev,Lvov:gauge} that the polarizability measured in
Compton scattering, $\overline{\alpha}$, can be written as 
$\overline{\alpha} = \alpha + \Delta\alpha$, where
\begin{equation}
\alpha = 2 \sum_{X \neq N} 
\frac{\mid \langle N \mid D_z \mid X \rangle \mid^2}{E_N-E_X},
\;\;\; \Delta\alpha = \frac{e^2 \langle r^2 \rangle_E}{3M}
+ \frac{e^2(\kappa^2+1)}{4 M^3} .
\label{alphas}
\end{equation}
Here $|N\rangle$ is the nucleon state, and $|X\rangle$ is any
intermediate excited state connected by the electric dipole operator $D_z$.
The first term in the expression for $\Delta\alpha$ is the so called
recoil correction, which involves the charge, $e$, the mass, $M$, 
and the mean squared charge radius of the particle, and
the second term is the Schwinger scattering term for a particle
with anomalous magnetic moment $\kappa$.
Our goal is to calculate the dispersive contribution $\alpha$ in 
Eq.~(\ref{alphas}).
Note that according to Eq.~(\ref{alphas}), 
this is equivalent to calculating the coefficient of the second-order
energy shift of the nucleon in the constant electric field.

First, we very briefly review the NJL model \cite{Meissner89,Meissner91}.
Our expressions are written in the Euclidean space-time.
In the leading-$N_c$ treatment the effective action results from
taking into account one-quark loops. It contains the quark part, 
$S_{\rm eff}^F$, the 
chirally invariant mesonic part, $S_{\rm eff}^M$,
and the chiral symmetry breaking part, $S_{\rm eff}^{\rm br}$, 
which provides the pion its mass
\begin{equation} \label{action}
S_{\rm eff}=S_{\rm eff}^F+S_{\rm eff}^M+S_{\rm eff}^{\rm br} \\
\end{equation}
\begin{eqnarray}
\label{action:F}
S_{\rm eff}^F = -{\rm Tr\, ln_\Lambda}(i \slashchar \partial
 - M U^{\gamma_5})-{\rm vac},\\
\label{action:M}
S_{\rm eff}^M={\mu^2 \over 2} \int d^4 x \;(\sigma^2+\mbf\pi^2)
-{\rm vac},\\
\label{action:br}
S_{\rm eff}^{\rm br}=-{m_0 \mu^2 \over g} \int d^4 x \; \sigma-{\rm vac}.
\end{eqnarray}
The trace is over the quark spinors.
The subscript $\Lambda$ reminds of the NJL cut-off (in this
paper we use the proper time regulator \cite{Schwinger51,Dyakonov86}),
$m_0$ is the current quark mass, and $M$ is the constituent quark mass.
The matrix $U^{\gamma_5}$ represents
the meson {\em hedgehog} profile in the non-linear representation
\begin{equation} \label{hedgehog}
U^{\gamma_5}(\mbf r) = \exp (i \gamma_5 \mbf\tau \cdot \mbf{\widehat r}
\Theta(r)) = {1\over f_\pi} (\sigma_h(\mbf r) +
i \gamma_5 \mbf\tau \cdot \mbf\pi_h(\mbf r)),
\end{equation}
where $\Theta(r)$ is the chiral angle, $\sigma_h(\mbf r) = \sigma_h(r)$ 
and
$\mbf\pi_h(\mbf r) = \widehat{\mbf r} \pi_h(r)$ are
the sigma and pion hedgehog profiles with
\mbox{$\sigma_h(r) = f_\pi \cos \Theta(r)$} and
\mbox{$\pi_h(r) = f_\pi \sin \Theta(r)$}, $\mbf{\widehat r}$ is the 
radial unit vector, and $\mbf\tau$ are the Pauli
isospin matrices. The  parameters of the model are fixed
by reproducing the experimental values of
the pion mass, $m_\pi$, and the pion decay constant, $f_\pi$, leaving the
quark-meson coupling constant $g$, or, equivalently, the constituent quark
mass $M = g f_\pi$ as the only free parameter \cite{Meissner91}.

Solitonic solutions to the NJL model are found by 
selfconsistently solving the coupled Euler-Lagrange equations 
for the profile function $\Theta$ and the Dirac spinors $q_{\rm n}$, which
satisfy the Dirac equation 
\begin{equation}
\label{Dirac:ham}
H q_{\rm n} 
= \left ( -i \mbf\nabla \cdot \mbf\alpha + 
\beta M U^{\gamma_5} \right ) q_{\rm n}
= \epsilon_{\rm n}  q_{\rm n}
\end{equation}

In the presence of the the electromagnetic field $A^\mu$, the minimal
substitution 
\mbox{$\partial^\mu \rightarrow \partial^\mu+ i e Q A^\mu$},
with the quark charge
\mbox{$Q=\left({1\over 2 N_c}+{1\over 2}\tau^3 \right)$},
leads to the following gauged effective action
\begin{equation} \label{action:gauged}
S_{\rm eff}^A = -{\rm Tr\, ln_\Lambda}(i \slashchar \partial -e Q
\slashchar A - M U^{\gamma_5})-{\rm vac}.
\end{equation}
The hedgehog solution brakes the
spin, $J$, and isospin, $I$, symmetries of the lagrangian, preserving the
grand spin $G=J+I$ symmetry.
To restore good quantum numbers we use the semiclassical ``cranking''
projection scheme \cite{ANW83,CB86}. 
Introducing the collective rotation matrix 
$R(t) = B \exp (i {{\mbf\Omega \cdot \mbf\tau}\over 2}t)$ in the usual way 
and transforming the quark fields and meson profiles
to the isorotating frame \cite{ANW83,CB86} we obtain
\begin{equation} \label{cranked}
\widetilde{S}^{A}_{\rm eff} = -{\rm Tr\, ln_\Lambda}(i \slashchar\partial
-e B^\dagger Q B
\slashchar A -i \beta {\mbf\Omega\cdot\mbf\tau \over 2}
-M U^{\gamma_5})-{\rm vac}.
\end{equation}
It is convenient to introduce the notation
$B^\dagger \tau^3 B = \mbox{\boldmath $c$}
\cdot \mbox{\boldmath $\tau$}$, where $c^a$ is an element
of a Wigner $D$-matrix, $c^a = D^1_{a 3}$ \cite{response}.
Then the quark electric charge in the isorotating frame can be written as
\begin{equation}
B^\dagger Q B = \frac{1}{2N_c} +
\frac{\mbf c \cdot \mbf\tau}{2}\,\cdot
\label{charge}
\end{equation}
The action
(\ref{cranked}) describes the system which is perturbed by {\em small}
perturbations: electromagnetic, and cranking. 
 From now on our method will follow very closely Ref.~\cite{response}.
The idea is the following: we have found the soliton, and now we 
want to calculate its linear response to external perturbations.
The perturbation operators
consist of products of two parts: {\em intrinsic}, which act on the 
internal structure of the state, and {\em collective},
which act on collective coordinates. For the cranking
perturbation, $i \beta {{\mbf\Omega\cdot\mbf\tau}\over 2}$, 
the collective operator
is \mbox{\boldmath $\Omega$}, and $i \beta \mbf\tau /2$ is the
intrinsic operator. Similarly, the electromagnetic perturbation has
a collective part (vector $\mbf{c}$ 
in Eq.~(\ref{charge})) and an intrinsic part.
Also, the nucleon state is a product of 
a collective and an intrinsic part \cite{ANW83,CB86,response}.
Formal perturbative expansion of Eq.~(\ref{cranked}) to $n$-th order
in small parameters involves expressions
of the general form 
\begin{eqnarray}
\kappa_{{i_1}, ... ,{i_n}} &=& 
 \langle coll|{\cal V}_{i_1}^{coll} 
{\cal V}_{i_2}^{coll} ... {\cal V}_{i_n}^{coll}|coll \rangle \nonumber \\
& & \int d^4 x_1 \; ... \; d^4 x_n \; 
\langle H | {\cal V}_{i_1}^{intr}(x_1) 
{G}(x_1,x_2) {\cal V}_{i_2}^{intr}(x_2) \nonumber \\
& & {G}(x_2,x_3) \; ... \; {\cal V}_{i_{n-1}}^{intr}(x_{n-1})
{G}(x_{n-1},x_{n}) 
{\cal V}_{i_n}^{intr}(x_n){G}(x_{n},x_{1}) | H \rangle, 
\label{eq:kappan}
\end{eqnarray}
where $|coll \rangle$ is the baryon collective wave function,
$| H \rangle$ is the intrinsic hedgehog state, 
${\cal V}_{i_k}$ is the perturbation of the 
$k^{th}$ type, and ${G}$ is the quark Feynman propagator in the 
background meson field, which
in the spectral representation has the form 
\begin{equation}
G(x,y) = \left ( 
\Theta(x_4-y_4) \sum_{n \in {\rm unocc.}} e^{-\epsilon_n (x_4-y_4)}
- \Theta(y_4-x_4) \sum_{n \in {\rm occ.}} e^{-\epsilon_n (x_4-y_4)}
\right ) q_{\rm n}(\mbf{x}) q^\dagger_{\rm n}(\mbf{y}) ,
\label{eq:G} 
\end{equation}
where {\footnotesize occ.} denotes all occupied states,
{\em i.e.} the valence as well as the Dirac sea states
and {\footnotesize unocc.} denotes the unoccupied positive energy states.

Examples of application of Eq.~(\ref{eq:kappan}) are given in
Ref.~\cite{response}.

\section{Leading-$N_c$ contributions to $\alpha$ }
\label{se:leading}
It was shown in Ref.~\cite{polar} that the leading contribution to 
the electric polarizability
of the nucleon scales as $N_c$.
Our first task is to obtain  this leading contribution  
for $\alpha$ in the NJL model. We denote it by 
$\alpha^{(0)}$. Since the cranking frequency $\mbf\Omega \sim 1/N_c$,
the leading-$N_c$ contribution is obtained by expanding the 
effective action (\ref{cranked}) to second order in
the electric field, and to zero order in $\mbf\Omega$.
 From Eq.~(\ref{charge}) we
notice, that for the leading $N_c$ 
contribution we have to keep only the isovector part of the charge,
since the isoscalar part is one power of $N_c$ down.

We now take 
$A^4 = Ez$, $\mbf{A} = 0$,  and expand Eq.~(\ref{action:gauged}) to 
second order in $E$ and zero order in $\mbf\Omega$.
We obtain the formula of the form (\ref{eq:kappan})
\begin{eqnarray} 
\label{alpha:new}
\alpha^{(0)} &=& - \frac{1}{T} \;e^2 \langle N \mid
{\rm Tr} \left ( {\mbf{c} \cdot \mbf\tau \over 2}\; x_3\; G(x,y)\;
{\mbf{c} \cdot \mbf\tau \over 2} \; y_3\; G(y,x) \right )
\mid N \rangle \nonumber \\
&=& \frac 12  {N_c}\;e^2 \langle N \mid c^a c^b \mid N\rangle
\sum_{m \in {\rm occ.} \atop n \in {\rm unocc.}}
{\langle m |\tau^a z | n \rangle\langle n|\tau^b z| m \rangle
\over \epsilon_n - \epsilon_m} ,
\end{eqnarray}
where $|N\rangle$ is the collective nucleon state, $T$ is
a large time interval, over which the time integrations are carried,
and $| n \rangle$ and $\epsilon_n$ denote the eigenstates and eigenvalues
of the Dirac hamiltonian (\ref{Dirac:ham}).
We can now split this formula into the valence and sea parts.
The sea contribution carries the cut-off function, and the vacuum
contribution has to be subtracted.
Finally, using 
\mbox{$\langle N\
\mid c^a c^b \mid N\rangle = \frac{1}{3} \delta^{ab}$},
we get the following formulas for the leading $N_c$ valence and sea
contributions to the electric polarizability of the nucleon
\begin{equation} \label{val}
\alpha^{(0)}_{\rm val} ={N_c\over 6}\;e^2 \sum_{n \ne {\rm val}}
{\langle {\rm val} |\tau^a z | n \rangle\langle n|\tau^a z| {\rm val} \rangle
\over \epsilon_n - \epsilon_{\rm val}} ;
\end{equation}
\begin{equation} \label{sea}
\left.
\alpha^{(0)}_{\rm sea}=
{N_c\over 6}\;e^2 \sum_{n \ne m} R(\epsilon_m,\epsilon_n,\Lambda)
{\langle m |\tau^a z | n \rangle\langle n|\tau^a z| m \rangle}
- \right. {\rm vac},
\end{equation}
where $R(\epsilon_m,\epsilon_n,\Lambda)$ is the
proper time regularization function
\begin{equation} \label{regular}
R(\epsilon_m,\epsilon_n,\Lambda)= -{1 \over 4 \sqrt \pi} \int_{\Lambda^{-2}}
^\infty {d\, s\over s^{3/2}}\left[{e^{-s \epsilon_m^2}- e^{-s \epsilon_n^2}
\over \epsilon_m^2 - \epsilon_n^2} +s {\epsilon_m e^{-s \epsilon_m^2}
+ \epsilon_n s e^{-s \epsilon_n^2}\over \epsilon_m + \epsilon_n} \right],
\end{equation}
with $\Lambda$ denoting the (proper time) cut-off parameter. 
The value of
$\Lambda$ is fit for any $M$ to reproduce the experimental value of
the pion coupling constant $f_\pi$ using the amplitude of the weak pion
decay \cite{Meissner91}.
The subtracted vacuum contribution in Eq.~(\ref{sea}) has the same
structure as the first term, but the free quark states and
eigenvalues are used. It may be immediately verified that both the valence
and sea contributions to $\alpha^{(0)}$ scale as $N_c$ \cite{polar}.

In the NJL model, predictions depending on the cut-off are somewhat
ambiguous. As we shall see, however, in our case $\alpha^{(0)}_{\rm sea}$
is dominated by the lowest term in the gradient
expansion, which
is independent of the cut-off.

The numerical results for the valence, $\alpha^{(0)}_{\rm val}$,
and Dirac sea, $\alpha^{(0)}_{\rm sea}$,
parts of $\alpha^{(0)}$ are
plotted in Fig.~1 as a function of the constituent quark mass $M$.
Both contributions decrease with increasing $M$, the valence curve
has much larger slope for smaller $M$ due to the fast increasing
energy gap between the valence orbital and the positive continuum.
For the optimum value \cite{Meissner91,Christo:ff} 
of $M = 420\ {\rm MeV}$,
we obtain \mbox{$\alpha^{(0)} = 16 \times 10^{-4}\ {\rm fm}^3$}.
The sea contributes to $\alpha^{(0)}$ 
65\%, and dominates over the
valence part, which gives 35\%.
Since the Dirac sea effects describe the physics of the pion cloud, 
this result is in qualitative agreement with the results of
other models \cite{Weiner85,polar,Golli:pol}.

We have also performed the gradient expansion of $\alpha^{(0)}_{\rm sea}$
up to first two terms following the scheme of 
Refs.~\cite{Aitchison,Arriola91,Schueren92}. This provides a link to the 
$\sigma$-model 
\cite{BirBan8485,polar} as well as an independent check for our numerical
procedure in the case of large size solitons. The leading-order term
in the gradient expansion of $\alpha^{(0)}_{\rm sea}$
gives us exactly the {\em seagull} term of the
$\sigma$-model
\begin{equation} \label{seagull}
\alpha_{\rm seagull}={8 \pi e^2\over 9}\int dr\, r^4
\left( \pi_h(r)\right)^2.
\end{equation}
Note that this seagull term arises in the NJL model from 
{\em purely dispersive} Dirac sea effects ({\em cf.} Eq.~(\ref{sea})).  
Recently, the inclusion of the seagull term in soliton models
such as the $\sigma$-model or the Skyrmion has been questioned
\cite{Lvov:gauge}. The NJL model provides a simple
interpretation of how Eq.~(\ref{seagull}) arises in the 
$\sigma$-model, and clears the doubts
of Ref.~\cite{Lvov:gauge}. 
Numerically the seagull 
term gives more than 80\% of the total sea contribution.

Looking at absolute numbers it may seem striking 
that our value of the seagull contribution,
$\sim 9 \times 10^{-4}\ {\rm fm}^3$, and the
$\sigma$-model \cite{polar} value, $\sim 30 \times 10^{-4}\ {\rm fm}^3$
differ so drastically.
The explanation lies in the large difference in the values of the axial
coupling constant at the leading $N_c$ level. Our model predicts 
$g_A^{(0)} \sim 0.75$, while in the $\sigma$-model 
$g_A^{(0)}=1.42$ \cite{CB86}.
The asymptotic pion tail, which gives the dominant
contribution in (\ref{seagull}), is proportional to $g_A^{(0)}$ and since
the pion field enters quadratically in (\ref{seagull}) the large
effect follows. We will come back to the question of $g_A$ in the
next section.

It is interesting to investigate in some greater detail
how well the gradient expansion works
for the electric polarizability.
In Fig.~2 we compare the results of
the gradient expansion with the full calculation 
of $\alpha^{(0)}_{\rm sea}$ for 
large-size solitons parameterized by the profile function involving
a size parameter $R$
\begin{equation} \label{profile}
\Theta(r)=-2\, {\rm atan} \left (
{R^2\over r^2} (1+m_\pi r) e^{-m_\pi r} \right).
\end{equation}
With increasing size of the soliton the result of the gradient 
expansion for $\alpha^{(0)}_{\rm sea}$ 
up to first two terms converges to the full result
very well. At $R=2$~fm the difference is only 2\%. Even the leading
order of the gradient expansion, $\alpha_{\rm seagull}$, shows good 
agreement with the full result, reproducing it up to 90\% at $R=2$~fm.

\section{Rotational corrections to $\alpha$ } 
\label{se:subleading}
Now we turn to the problem of the next-to-leading order 
corrections in the $1/N_c$ expansion of the electric
polarizability $\alpha$.
As already mentioned above, the 
value of $g_A$ in chiral soliton models is
crucial for the results for $\alpha$. 
The prediction for $g_A$ in the
NJL model (as well as in the Skyrme model) at the leading-$N_c$ 
level, denoted by $g_A^{(0)}$, turns out 
to be much too small, $\sim 0.7-0.8$ 
\cite{Meissner91:ga}, and no parameter
fits or extensions of the model were able to raise this value. 
Recently, Ref.~\cite{Wakamatsu93} proposed how to include rotational
$1/N_c$ corrections to $g_A$ in the NJL model. A correct and detailed
study of this effect has been presented in 
Refs.~\cite{Christo:ga,Michal:ga,Andree:ga}.
The nature of these corrections is very easy to understand, 
for instance 
from the point of view of the formalism of Ref.~\cite{response}.
At the leading-$N_c$ level the expression for $g_A$ involves 
a sum of matrix elements of the axial operator in the quark
single-particle states~\cite{Meissner91:ga}. However, in the presence
of cranking the hedgehog state is perturbed, and we have
a situation where one of the
perturbations is the axial vector current, and the other one is the cranking.
The effect discussed in Refs.~\cite{Wakamatsu93,Michal:ga,Christo:ga} is
associated with the noncommutativity of the collective parts of the
corresponding operators (the cranking
and the axial vector current operators in this case). An analogous
effect arises for the case of $\alpha$, as we discuss below. 

Since the cranking operator
scales as $1/N_c$ (it involves the cranking frequency
$\mbf\Omega \sim 1/N_c$) the correction is suppressed by one
power of $N_c$. This correction to
the axial vector coupling constant, denoted by $g_A^{(1)}$,
has the valence part and the sea part. The numerical value for 
$g_A^{(1)}$ is about $0.5$ \cite{Wakamatsu93,Christo:ga,Michal:ga}, 
and raises the total 
value of $g_A$ to a comfortable number.
Of course, the described $1/N_c$ corrections are not the only possible ones,
but they seem to be particularly important. 

We have explained before that the small value of $g_A^{(0)}$ reduces the value
of the seagull contribution to the polarizability, which is a desired effect.
Since roatational $1/N_c$ corrections 
raise significantly the value of $g_A$, we have
also to examine how analogous effects 
influence the electric polarizability. 
The $1/N_c$ 
rotational corrections to $\alpha$ are obtained by
expanding the effective action~(\ref{cranked}) 
to second order in $E$ and first order in $\mbf\Omega$.
Thus, we have formally the case of the third-order perturbation theory.
We get expression of the generic form (\ref{eq:kappan}), with all
possible time-orderings of collective operators.

The relevant collective matrix elements for the present 
case have the form
\begin{equation}
\langle N \mid c^a \Omega^b c^c \mid N \rangle 
= \frac{1}{3 {\cal J}} \epsilon^{abc} ,
\label{eq:coleps}
\end{equation}
where $\cal J$ is the moment of inertia. 
Finally, after straightforward algebra, we obtain
for the valence contribution
\begin{eqnarray} \label{omega1val}
\lefteqn{
\alpha^{(1)}_{\rm val} = 
- {{i N_c e^2 \epsilon^{abc}} \over {12 {\cal J}}}}\\  
& & \times \left [ \sum_{n>0, n \neq {\rm val} \atop m>0, m \neq {\rm val}}
{\langle {\rm val} |\tau^a z | n \rangle\langle n|\tau^b| m \rangle
\langle m |\tau^c z | {\rm val} \rangle
\over (\epsilon_n - \epsilon_{\rm val}) (\epsilon_m - \epsilon_{\rm val})} 
- \sum_{n<0 \atop m<0} 
{\langle {\rm val} |\tau^a z | n \rangle\langle n|\tau^b| m \rangle
\langle m |\tau^c z | {\rm val} \rangle
\over (\epsilon_n - \epsilon_{\rm val}) (\epsilon_m - \epsilon_{\rm val})}
\right. \nonumber \\
& & - \left. 2  \sum_{n>0, n \neq {\rm val} \atop m<0} 
{\langle {\rm val} |\tau^a z | n \rangle\langle n|\tau^b z| m \rangle
\langle m |\tau^c | {\rm val} \rangle
\over (\epsilon_n - \epsilon_{\rm val}) (\epsilon_n - \epsilon_m)} 
+2 \sum_{n<0 \atop m>0}
{\langle {\rm val} |\tau^a z | n \rangle\langle n|\tau^b z| m \rangle
\langle m |\tau^c| {\rm val} \rangle
\over (\epsilon_n - \epsilon_{\rm val}) (\epsilon_n - \epsilon_m)}
\right ] \nonumber ,
\end{eqnarray} 
and for the sea contribution 
\begin{eqnarray} \label{omega1sea}
\lefteqn{
\alpha^{(1)}_{\rm sea} = 
- {{i N_c e^2 \epsilon^{abc}} \over {12 {\cal J}}}}\\  
& & \times \left [ \sum_{k<0 \atop m,n>0}
{\langle k |\tau^a z | n \rangle\langle n|\tau^b| m \rangle
\langle m |\tau^c z | k \rangle
\over (\epsilon_k - \epsilon_m) (\epsilon_k - \epsilon_n)} 
+ \sum_{k>0 \atop m,n<0}
{\langle k |\tau^a z | n \rangle\langle n|\tau^b| m \rangle
\langle m |\tau^c z | k \rangle
\over (\epsilon_k - \epsilon_m) (\epsilon_k - \epsilon_n)}
\right ] - {\rm vac} . \nonumber 
\end{eqnarray} 
In the above expressions notation $n<0$ means the level with
$\epsilon_n <0$, {\em etc.}
Recalling that ${\cal J} \sim N_c$, it can be verified that $\alpha^{(1)}$
is of order $N_c^0$, while the leading contribution to $\alpha^{(0)}$ was
of order $N_c$.

The numerical value for $\alpha^{(1)}_{\rm val}$ is obtained directly
from Eq.~(\ref{omega1val}). It involves a double summation over
quark levels, and computer time involved is similar to the
calculation of, say,  the moment of inertia.
Our results are shown in Fig.~1.
The value of $\alpha^{(1)}_{\rm val}$ is about 
\mbox{$3 \times 10^{-4}\ {\rm fm}^3$} for $M=420\ {\rm MeV}$,
so the effect is about 60\% of the leading-$N_c$ valence contribution
$\alpha^{(0)}_{\rm val}$. The decrease of 
$\alpha^{(1)}_{\rm val}$ with increasing $M$ is
similar to the behavior of the leading-$N_c$ valence contribution
$\alpha^{(0)}_{\rm val}$.

Now we turn to the sea contribution $\alpha^{(1)}_{\rm sea}$.
This contribution in principle
may involve a cut-off, similar to other 
observables calculated in the NJL model.
However, since Eq.~(\ref{omega1sea}) involves a 
computer-time-consuming triple sum over quark states,
we approximate $\alpha^{(1)}_{\rm sea}$ by the leading term
in the gradient expansion.
As a bonus, we get the nice feature that the result is
independent of the cut-off, exactly as in the case of the
leading-$N_c$ contribution $\alpha_{\rm seagull}$.
In the present case the gradient expansion  technique involves some 
tricks \cite{Pasha}, 
so we sketch the method in some greater detail.  
The first step is to rewrite Eq.~(\ref{omega1sea}) in a form
involving unconstrained sums, {\em i.e.} sums over all 
quark levels. This can be achieved using the identity true for any
real numbers $a$, $b$ and $c$
\begin{eqnarray}
&&\int \frac{d \omega_1 d \omega_2 d \omega_3}{(2 \pi)^3}
\frac{1}{\omega_1 - ia}\;\frac{1}{\omega_1 - ib}\;\frac{1}{\omega_1 - ic}
\;\frac{1}{\omega_1 - \omega_2 - i \eta}\;
\frac{1}{\omega_1 - \omega_3 - i \eta}\nonumber\\
&& \;\;\; = -i \frac{1}{a-b}\;\frac{1}{a-c}\;\Theta(-a)\;\Theta(b)
\;\Theta(c) .
\label{eq:ident}
\end{eqnarray}
With help of Eq.~(\ref{eq:ident}) we can replace the constrained sums
in Eq.~(\ref{omega1sea}) by sums over all indices. For instance
\begin{eqnarray}
&&\sum_{k<0 \atop m,n>0} 
{\langle k |\tau^a z | n \rangle\langle n|\tau^b| m \rangle
\langle m |\tau^c z | k \rangle
\over (\epsilon_k - \epsilon_m) (\epsilon_k - \epsilon_n)} \nonumber \\
&&\;\;\; = i \sum_{k,m,n}
\langle k |\tau^a z | n \rangle\langle n|\tau^b| m \rangle
\langle m |\tau^c z | k \rangle
\int \frac{d \omega_1 d \omega_2 d \omega_3}{(2 \pi)^3}
\frac{1}{\omega_1 - i \epsilon_k}\;
\frac{1}{\omega_1 - i \epsilon_m}\;
\frac{1}{\omega_1 - i \epsilon_n} \nonumber \\
&&\;\;\;\times \left ( P \frac{1}{\omega_1 - \omega_2} +
i \pi \delta(\omega_1 - \omega_2) \right )
\left ( P \frac{1}{\omega_1 - \omega_3} + 
i \pi \delta(\omega_1 - \omega_3) \right ) , 
\label{eq:unconstr}
\end{eqnarray}
where $P$ denotes the principal value integral.
Introducing the trace over the quark spinors
$Tr$ we can now rewrite Eq.~(\ref{omega1sea})
as 
\begin{eqnarray}
\label{eq:gr0}
\alpha^{(1)}_{\rm sea} &=& \frac{i \epsilon^{abc}\; e^2}{12 {\cal J}}
\int \frac{d \omega_1 d \omega_2}{(2 \pi)^2} P \frac{1}
{\omega_1 - \omega_2} \\
&&\hspace*{-15pt}\times {\rm Tr} \left [
\tau^a z \frac{1}{\omega_1 - i H}
\tau^b   \frac{1}{\omega_2 - i H}
\tau^c z \frac{1}{\omega_1 - i H} +
\tau^a z \frac{1}{\omega_1 - i H}
\tau^b   \frac{1}{\omega_1 - i H}
\tau^c z \frac{1}{\omega_2 - i H} \right ] -{\rm vac}. \nonumber
\end{eqnarray}
Standard gradient expansion techniques may now be used to expand
the functional trace in Eq.~(\ref{eq:gr0}) \cite{Aitchison}.
The lowest order term (denoted by ${}_{\rm lowest}$),
contains no gradients, and we obtain
\begin{eqnarray}
&&\alpha^{(1)}_{\rm sea, lowest} 
= - \frac{32 N_c M^2}{3 {\cal J} f_\pi^2}
\int \frac{d^3 k}{(2 \pi)^3} 
\int \frac{d \omega_1 d \omega_2}{(2 \pi)^2}\;P \frac{1}{\omega_1 - \omega_2}
\frac{(2 \omega_1 + \omega_2)}{(\omega_1^2+k^2+M^2)^2 
(\omega_2^2 + k^2 + M^2)} \nonumber \\
&&\;\;\;\times \int d^3 r \;z^2 (\pi_h (r))^2 \nonumber \\
&&\;\;\; = \frac{N_c M}{16 \pi {\cal J} f_\pi^2} \;\alpha_{\rm seagull} ,
\label{eq:grad1}
\end{eqnarray} 
where $\alpha_{\rm seagull}$ is given in Eq.~(\ref{seagull}).

For $M=420\ {\rm MeV}$ the moment of inertia has the value
${\cal J} = 1.1\ {\rm fm}$, and we obtain
for the proportionality factor in the last equality in 
Eq.~(\ref{eq:grad1}) the value $0.54$. The numerical results for
$\alpha^{(1)}_{\rm sea, lowest}$ are plotted in Fig.~1. The
contribution increases with increasing constituent quark mass $M$. This is
due to the decrease of the moment of inertia with $M$. 
Guided by our experience from gradient-expanding the leading-$N_c$
contribution (Sec.~\ref{se:leading}), 
we hope that also in the present case the lowest term
in the gradient expansion almost saturates the value of 
$\alpha^{(1)}_{\rm sea}$.

At the end of this section we would like to present an interesting
observation. The formula analogous to Eq.~(\ref{eq:grad1}) for the case 
of $g_A$ has the form \cite{Andree:ga}
\begin{equation}
g_{A, {\rm sea, lowest}}^{(1)} = \frac{N_c M}{32 \pi {\cal J} f_\pi^2} 
\;g^{(0)}_{A, {\rm sea, lowest}} .
\label{eq:gagrad1}
\end{equation} 
Note that the coefficient between the $1/N_c$ correction and the leading-$N_c$
term in Eq.~(\ref{eq:gagrad1}) is {\em exactly twice} the analogous 
coefficient in Eq.~(\ref{eq:grad1}). This is not a coincidence.
In the chiral limit, $m_\pi \to 0$, the sea contribution to $\alpha$
is divergent as $1/m_\pi$ \cite{BKM91,polar,chihog}. This divergent
piece is proportional to $g_A^2$, and this result is a general, model
independent feature of the chiral limit. Therefore, one has
$\alpha_{\rm sea} = \frac{\gamma}{m_\pi} g_A^2$, where $\gamma$ is a numerical
constant. Hence, if in a model
one corrects $g_A$ by including $1/N_c$ terms, analogous correction
should come out in the calculation of $1/N_c$ corrections to $\alpha$.
Consider the formal case of solitons with a very large size $R$.
In that limit, $g_A$ consists only of the sea contribution, which
in turn is saturated by the lowest term in the gradient expansion:
$g_{A} = g_{A, {\rm sea, lowest}}$.  
Using Eq.~(\ref{eq:gagrad1})
and the general considerations discussed above, we obtain
in the double limit of $m_\pi \to 0$ and $R \to \infty$ the following
result
\begin{eqnarray}
\label{eq:chirlim}
\alpha_{\rm sea, lowest} &=&
\frac{\gamma}{m_\pi} g_A^2 =
\frac{\gamma}{m_\pi} 
\left (g_{A, {\rm sea, lowest}}^{(0)} + 
g_{A, {\rm sea, lowest}}^{(1)} \right )^2 + {\cal O}(1/N_c^2) \nonumber \\
&=& \alpha_{\rm seagull} +
\frac{N_c M}{16 \pi {\cal J} f_\pi^2} 
\alpha_{\rm seagull} 
+ {\cal O}(1/N_c^2) .
\end{eqnarray}
The first term in the last line of is our 
$\alpha^{(0)}_{\rm sea, lowest}$, the
second term is $\alpha^{(1)}_{\rm sea,lowest}$, 
with exactly the same factor as in Eq.~(\ref{eq:grad1}).
This shows that our method of including rotational $1/N_c$
corrections is consistent with chiral physics.  

Analogous consistency conditions have to hold for other observables which
behave as $g_A^2 / m_\pi$, {\em e.g.} the isovector magnetic mean squared
radius of the nucleon.

Another digression concerns the structure of the rotational $1/N_c$
corrections. Collective operators which arise in evaluating corrections to
$g_A$ or the isovector magnetic moment have the same structure 
as collective operators entering at leading-$N_c$ level. Formally, 
this is because $[J^a,D^1_{bc}] = i \epsilon^{abd} D^1_{dc}$.
Thus, ratios of these observables (also for non-diagonal
matrix elements) for various members of the flavor
SU(2) multiplet ({\em i.e.} for $N$ and $\Delta$) 
are preserved at the $1/N_c$ level, and
obtain corrections starting at the $1/N_c^2$ level. This complies to
the consistency condition of Ref.~\cite{DashenM}.

\section{Effects of the $\Delta$-$N$ mass splitting}
\label{se:splitting}
Before comparing our results to data, we note that there is another very
important effect which has not been included, namely the 
$N$-$\Delta$ mass splitting. As discussed in Ref.~\cite{polar,chihog},
in the limit $N_c \to \infty$ our perturbative approach has the nucleon
and the $\Delta$ degenerate in mass. At the
hadronic level, the seagull term results from diagrams with 
intermediate pion-nucleon or pion-$\Delta$ states. Because of the 
$N$-$\Delta$ degeneracy, the contribution of the 
pion-$\Delta$ states is strongly overestimated in the hedgehog treatment. 
One may crudely include the
effect of $N$-$\Delta$ mass splitting by 
reducing the seagull contribution  by about 30\%
\cite{polar,chihog}. A consistent treatment of this effect would
require the use of a better collective quantization method,
but this does not seem feasible for the NJL model.
As shown in Section~\ref{se:leading} more than 80\% of the sea
contribution $\alpha^{(0)}_{\rm sea}$ is carried by the seagull term.
Therefore, as suggested in Refs.~\cite{polar,chihog}, we simply multiply
the sea contribution by a factor of $2/3$ to mimic the effect
of $N$-$\Delta$ mass splitting.

Since $\alpha^{(1)}_{\rm sea, lowest}$ 
is proportional to $\alpha_{\rm seagull}$,
we expect that it obtains analogous corrections from $\Delta$-$N$ splitting
as the leading-$N_c$ part. Therefore we also multiply
it by a factor of $2/3$ before comparing to data.

\section{Final result and conclusion} \label{se:conclusion} 
Our numerical results are collected in Figs.~1,3. They are plotted as
a function of the constituent quark mass $M$. Figure~1  shows the sea
and valence contributions to the leading-$N_c$ nucleon electric 
polarizability $\alpha^{(0)}$ as well as to the rotational $1/N_c$
correction $\alpha^{(1)}$. The total polarizability is presented on
Fig.~3 compared to the experimental value. The dashed line corresponds
to $\alpha' =  \alpha^{(0)}_{\rm sea}+\alpha^{(0)}_{\rm val}
+ \alpha^{(1)}_{\rm sea,lowest}+\alpha^{(1)}_{\rm val}$. The full
line shows the total $\alpha$, crudely corrected for the effects of the
$N$-$\Delta$ mass splitting by multiplying the sea contributions
by a factor of $2/3$ (Sec.~\ref{se:splitting}):
$\alpha = \frac23 \alpha^{(0)}_{\rm sea}+\alpha^{(0)}_{\rm val}
+ \frac23 \alpha^{(1)}_{\rm sea,lowest}+\alpha^{(1)}_{\rm val}$.
For values of $M$ in the physically relevant range 400--450~MeV the
changes of $\alpha$ with  $M$
are small. Choosing for $M=420$~MeV
(value which gives good agreement for
other nucleon observables \cite{Meissner91,Christo:ff}) we obtain for the total
nucleon electric polarizability the value
$\alpha=19\times 10^{-4}\ \rm fm^3$, with
$\frac23 \alpha^{(0)}_{\rm sea}$, $\alpha^{(0)}_{\rm val}$,
$\frac23 \alpha^{(1)}_{\rm sea,lowest}$,
and $\alpha^{(1)}_{\rm val}$
contributing 36\%, 29\%, 17\%, and 17\% respectively. Our result
is still too large compared to the experimental number
$\alpha_{\rm exp}=9.6 \pm 1.8 \pm 2.2
\times 10^{-4}\ \rm fm^3$ \cite{Federspiel,Zieger,Schmiedmayer}. 
Keeping in mind the approximate nature of the model and techniques used,
our model prediction is within expectations. 

Note, that in our calculation, after including the rotational $1/N_c$
effects, we have the correct value of $g_A$. This is not the case of 
other models \cite{Chemtob87,Scoccola90,Schwesinger:pol}, 
and apparent agreement of $\alpha$ with experiment
in these models results from having too low $g_A$.

The main points of this paper can be summarized as follows:

1) Dirac sea effects dominate the electric polarizability of the
nucleon. In the gradient expansion
the sea effects are dominated by the seagull contribution. This
explains how this term arises in other models which do not have
the explicit Dirac sea. 

2) We have calculated rotational $1/N_c$ corrections, and
found that they are important and of expected size.

3) We have crudely estimated the $N$-$\Delta$ splitting
effects.

4) With all above effects included, numerical predictions of  
the NJL model are better than in other soliton models (note that
we have the correct value of $g_A$!) but are still about a factor
of $2$ too large compared to experiment. This indicates that 
the NJL soliton is too ``soft'' and other effects are still
needed in order to describe the polarizability properly.

5) Calculation of the magnetic polarizability, or the proton-neutron
splitting of the electric
polarizability would require a much greater effort in the NJL model
than in the $\sigma$-model calculation of Ref.~\cite{BBC91a,polar}. This
is because of the necessity to subtract zero-modes. 

\vspace{15mm}
 
The authors are  grateful to Pavel Pobylitsa for helpful comments. 
This work has been supported in part by 
the Polish State Committee for Scientific Research,
grants 2.0204.91.01 and 2.0091.91.01, and by the Bulgarian National
Science Foundation, contract $\Phi$-32.

\eject

\eject

\section*{Figure captions}
\vspace{10mm}

\noindent 
Fig.\ 1. 

\noindent
The electric polarizability of the nucleon, $\alpha$,
as a function of the constituent quark mass $M$.
Leading $N_c$ Dirac sea contribution $\alpha^{(0)}_{\rm sea}$ 
(dashed line); leading $N_c$ valence
contribution $\alpha^{(0)}_{\rm val}$ (dash-dotted line);
rotational $1/N_c$ correction to the sea contribution 
$\alpha^{(1)}_{\rm sea, lowest}$ (dotted line) and to the valence one
$\alpha^{(1)}_{\rm val}$ (dash-dot-dot).

\vspace{10mm}

\noindent 
Fig.\ 2.

\noindent
The ratio of the gradient expanded electric nucleon polarizability
to the full leading-$N_c$ 
$\alpha^{(0)} =\alpha^{(0)}_{\rm sea} + \alpha^{(0)}_{\rm val}$ for different
soliton sizes $R$. The dash-dotted line represents the leading order
in the gradient expansion $\alpha_{\rm seagull}$ and the dashed
line the seagull plus the next term in the gradient expanded $\alpha$.
 
\vspace{10mm}
 
\noindent 
Fig.\ 3.

\noindent
The total electric polarizability of the nucleon, $\alpha$,
as a function of the constituent quark mass $M$.
The dashed line shows the full $\alpha' =
\alpha^{(0)}_{\rm sea} + \alpha^{(0)}_{\rm val} + 
\alpha^{(1)}_{\rm sea,lowest} + \alpha_{\rm val}^{(1)}$
and the full line  $\alpha = {2 \over 3}
\alpha^{(0)}_{\rm sea} + \alpha^{(0)}_{\rm val} + 
{2 \over 3} \alpha^{(1)}_{\rm sea,leading} + \alpha_{\rm val}^{(1)}$,
where the factor ${2 \over 3}$ in front of $\alpha^{(i)}_{\rm sea}$
crudely corrects for the overestimation of the $\Delta$ contribution
(Sec.~\ref{se:splitting}).

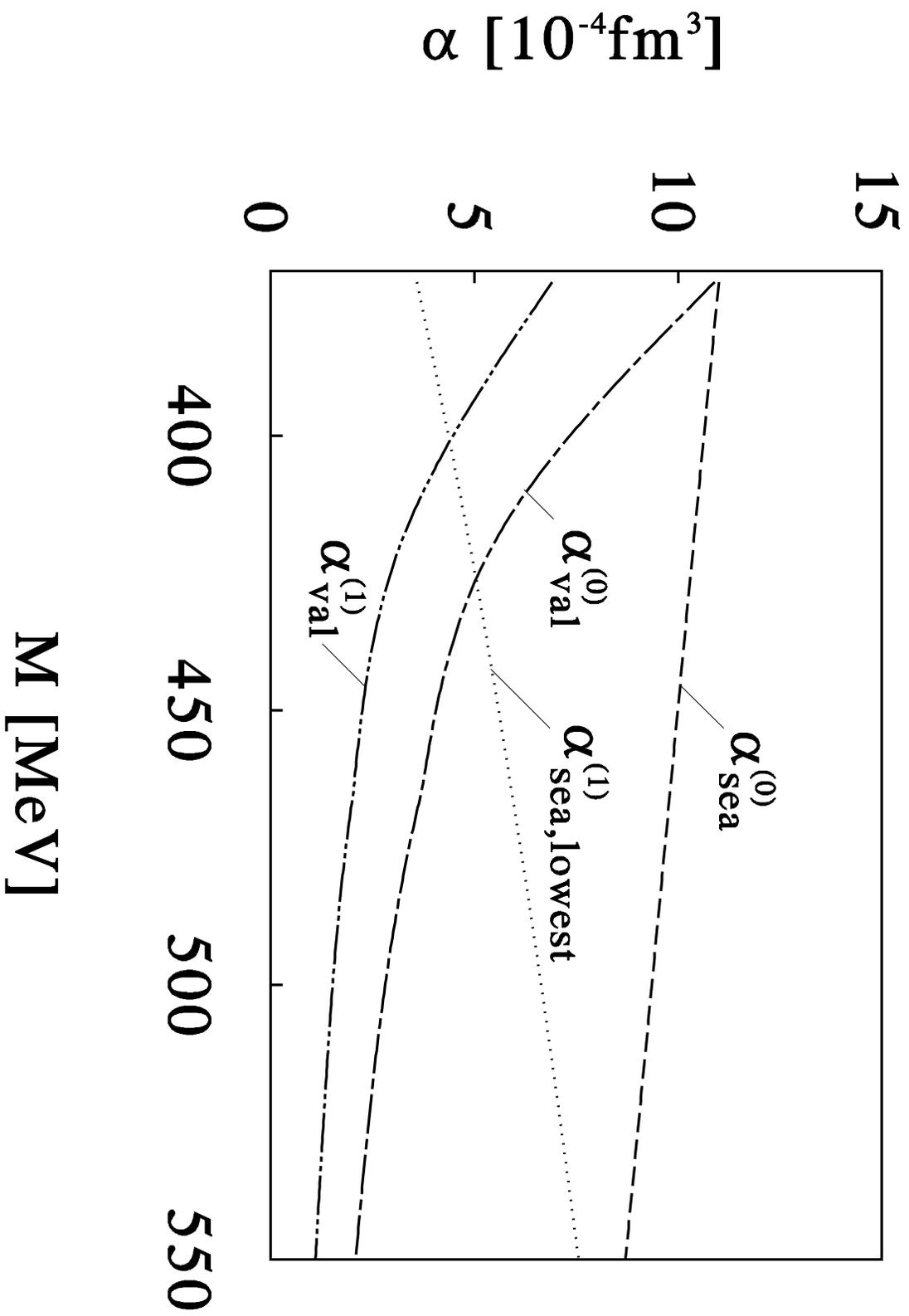
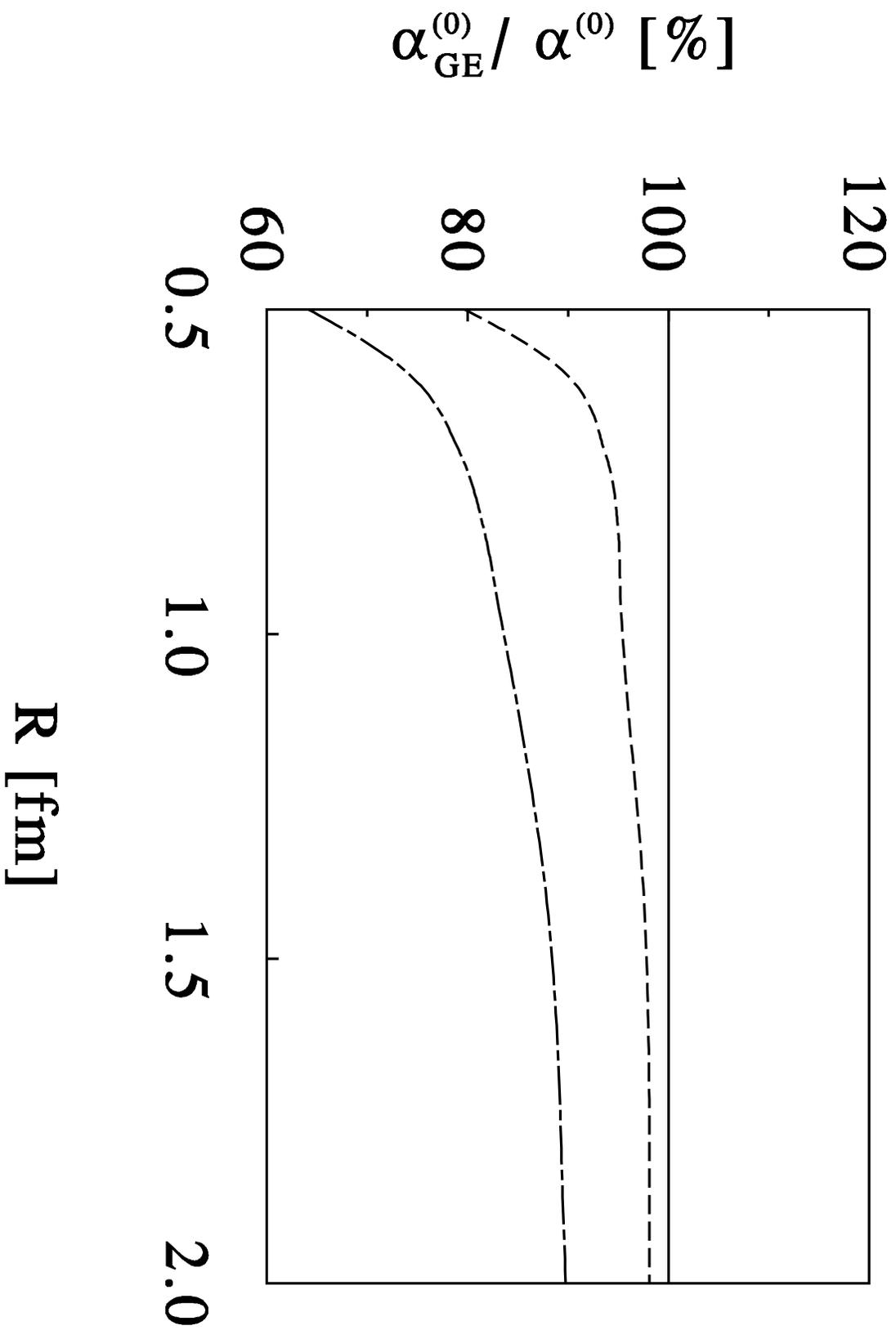
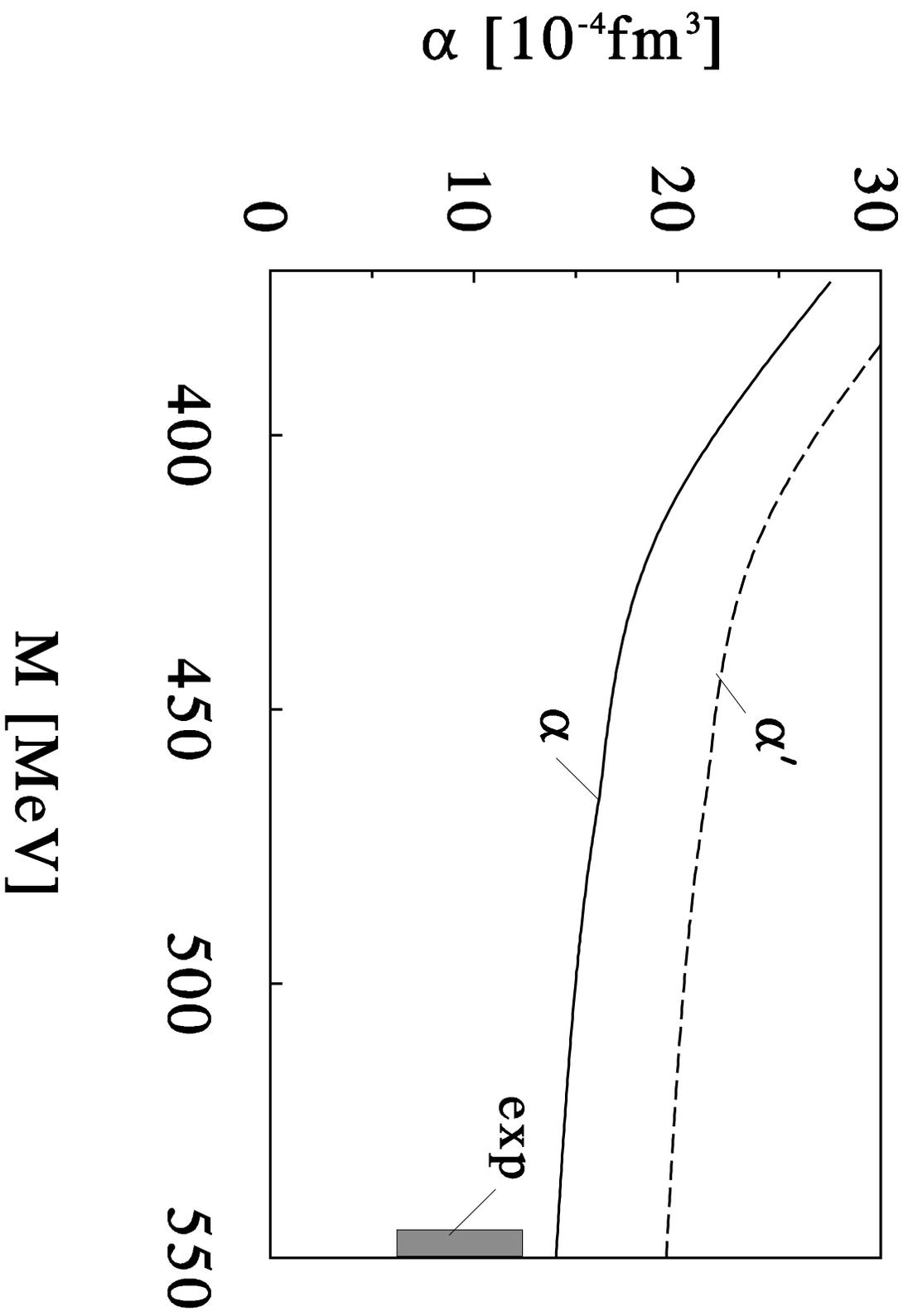
\end{document}